\documentclass[prl,aps,twocolumn,preprintnumbers,amsmath,amssymb,superscriptaddress,showpacs]{revtex4}

\usepackage{graphicx}
\usepackage{dcolumn}
\usepackage{bm}
\usepackage{epstopdf}

\newcount\driver
\newcount\bozza

scaled\magstep1                  
scaled\magstep1 
 
scaled\magstep1                 \font\indbf=cmbx10 scaled\magstep2

scaled \magstep2

{\count255=\time\divide\count255 by 60
\xdef\hourmin{\number\count255}
        \multiply\count255 by-60\advance\count255 by\time
   \xdef\hourmin{\hourmin:\ifnum\count255<10 0\fi\the\count255}}

     \let\g=\gamma     \let\d=\delta     \let\e=\varepsilon
  \let\h=\eta      \let\k=\kappa     \let\l=\lambda
                          \let\r=\rho
\let\s=\sigma          \let\ph=\varphi   
\let\ps=\psi   \let\o=\omega     
        \let\L=\Lambda

\def\xx{{\bf x}}
  
\def\yy{{\bf y}}\def\nn{{\bf n}}

       \def\oo{{\underline \omega}}
\def\ee{{\underline \varepsilon}}

\let\==\equiv

\let\io=\infty
\let\0=\noindent

\def\*{{\hfill\break\null\hfill\break}}

\def\tilde#1{{\widetilde #1}}

\def\tende#1{\,\vtop{\ialign{##\crcr\rightarrowfill\crcr
             \noalign{\kern-1pt\nointerlineskip}
             \hskip3.pt${\scriptstyle #1}$\hskip3.pt\crcr}}\,}
\def\otto{\,{\kern-1.truept\leftarrow\kern-5.truept\to\kern-1.truept}\,}

\def\wh#1{\widehat{#1}}
\def\hat#1{\wh{#1}}
\def\sqt[#1]#2{\root #1\of {#2}}

\def\bp{{\bar \ps}}

\def\T#1{{#1_{\kern-3pt\lower7pt\hbox{$\widetilde{}$}}\kern3pt}}
\def\VVV#1{{\underline #1}_{\kern-3pt
\lower7pt\hbox{$\widetilde{}$}}\kern3pt\,}
\def\W#1{#1_{\kern-3pt\lower7.5pt\hbox{$\widetilde{}$}}\kern2pt\,}

\def\indica{\leaders \hbox to 0.5cm{\hss.\hss}\hfill}
\def\guida{\leaders\hbox to 1em{\hss.\hss}\hfill}
\mathchardef\oo= "0521

\def\xx{{\bf x}}
\def\yy{{\bf y}}\def\nn{{\bf n}}

\def\oo{{\underline \omega}}

\def\qed{\raise1pt\hbox{\vrule height5pt width5pt depth0pt}}
 
  \def\bp{{\bar p}} 

\def\indic{\hbox{\raise-2pt \hbox{\indbf 1}}}

\def\sech{{\rm sech}}

\def\ins#1#2#3{\vbox to0pt{\kern-#2 \hbox{\kern#1 #3}\vss}\nointerlineskip}

\newdimen\xshift \newdimen\xwidth \newdimen\yshift

\def\insertplot#1#2#3#4#5#6{%
\xwidth=#1pt \xshift=\hsize \advance\xshift by-\xwidth \divide\xshift by 2%
\begin{figure}[ht]
\vspace{#2pt} \hspace{\xshift}
\begin{minipage}{#1pt}
#3 \ifnum\driver=1 \griglia=#6
\ifnum\griglia=1 \openout13=griglia.ps \write13{gsave .2
setlinewidth} \write13{0 10 #1 {dup 0 moveto #2 lineto } for}
\write13{0 10 #2 {dup 0 exch moveto #1 exch lineto } for}
\write13{stroke} \write13{.5 setlinewidth} \write13{0 50 #1 {dup 0
moveto #2 lineto } for} \write13{0 50 #2 {dup 0 exch moveto #1
exch lineto } for} \write13{stroke grestore} \closeout13
\includegraphics{griglia.ps} \fi
\includegraphics{#4.ps}\fi%
\ifnum\driver=2 \fi
\end{minipage}
\caption{#5}
\end{figure}
}

\newdimen\shift \shift=-1.5truecm
\def\lb#1{%
\ifnum\bozza=1
\label{#1}\rlap{\hbox{\hskip\shift$\scriptstyle#1$}}
\else\label{#1} \fi}

\def\be{\begin{equation}}
\def\ee{\end{equation}}
\def\bea{\begin{eqnarray}}\def\eea{\end{eqnarray}}
\def\bean{\begin{eqnarray*}}\def\eean{\end{eqnarray*}}
\def\bfr{\begin{flushright}}\def\efr{\end{flushright}}
\def\bc{\begin{center}}\def\ec{\end{center}}
\def\bal{\begin{align}}\def\eal{\end{align}}
\def\ba#1{\begin{array}{#1}} \def\ea{\end{array}}
\def\bd{\begin{description}}\def\ed{\end{description}}
\def\bv{\begin{verbatim}}\def\ev{\end{verbatim}}
\def\nn{\nonumber}
\def\Halmos{\hfill\vrule height10pt width4pt depth2pt \par\hbox to \hsize{}}
\def\pref#1{(\ref{#1})}

\def\ins#1#2#3{\vbox to0pt{\kern-#2 \hbox{\kern#1 #3}\vss}\nointerlineskip}

\newdimen\xshift \newdimen\xwidth \newdimen\yshift
\newcount\griglia

\def\insertplot#1#2#3#4#5#6{%
\xwidth=#1pt \xshift=\hsize \advance\xshift by-\xwidth \divide\xshift by 2%
\begin{figure}[ht]
\vspace{#2pt} \hspace{\xshift}
\begin{minipage}{#1pt}
#3 \ifnum\driver=1 \griglia=#6
\ifnum\griglia=1 \openout13=griglia.ps \write13{gsave .2
setlinewidth} \write13{0 10 #1 {dup 0 moveto #2 lineto } for}
\write13{0 10 #2 {dup 0 exch moveto #1 exch lineto } for}
\write13{stroke} \write13{.5 setlinewidth} \write13{0 50 #1 {dup 0
moveto #2 lineto } for} \write13{0 50 #2 {dup 0 exch moveto #1
exch lineto } for} \write13{stroke grestore} \closeout13
\includegraphics{griglia.ps} \fi
\includegraphics{#4.ps}\fi%
\ifnum\driver=2 \fi
\end{minipage}
\caption{#5}
\end{figure}
}

\newdimen\shift \shift=-1.5truecm
\def\lb#1{%
\label{#1}\rlap{\hbox{\hskip\shift$\scriptstyle#1$}}
\else\label{#1} \fi}

\def\be{\begin{equation}}
\def\ee{\end{equation}}
\def\bea{\begin{eqnarray}}\def\eea{\end{eqnarray}}
\def\bean{\begin{eqnarray*}}\def\eean{\end{eqnarray*}}
\def\bfr{\begin{flushright}}\def\efr{\end{flushright}}
\def\bc{\begin{center}}\def\ec{\end{center}}
\def\bal{\begin{align}}\def\eal{\end{align}}
\def\ba#1{\begin{array}{#1}} \def\ea{\end{array}}
\def\bd{\begin{description}}\def\ed{\end{description}}
\def\bv{\begin{verbatim}}\def\ev{\end{verbatim}}
\def\nn{\nonumber}
\def\Halmos{\hfill\vrule height10pt width4pt depth2pt \par\hbox to \hsize{}}
\def\pref#1{(\ref{#1})}

\usepackage{amsmath}
\usepackage{amsfonts}
\usepackage{amssymb}
\usepackage{epstopdf}

scaled\magstep1

   \let\g=\gamma  \let\d=\delta
\let\e=\varepsilon
  \let\h=\eta    \let\k=\kappa \let\l=\lambda
                 \let\r=\rho
\let\s=\sigma     \let\ph=\varphi
\let\ps=\Psi   \let\o=\omega
   \let\L=\Lambda

 \def\HHH{{\cal H}}

 \def\xx{{\bf x}} \def\yy{{\bf y}}

\def\nn{\nonumber}

\def\\{\hfill\break}
\def\={:=}
\let\io=\infty
\let\0=\noindent

\def\tende#1{\,\vtop{\ialign{##\crcr\rightarrowfill\crcr\noalign{\kern-1pt
    \nointerlineskip} \hskip3.pt${\scriptstyle #1}$\hskip3.pt\crcr}}\,}
\def\otto{\,{\kern-1.truept\leftarrow\kern-5.truept\to\kern-1.truept}\,}

\def\wh{\widehat}
\def\to{\rightarrow}

\def\qed{\hfill\raise1pt\hbox{\vrule height5pt width5pt depth0pt}}

\def\be{\begin{equation}}
\def\ee{\end{equation}}
\def\bp{\begin{pmatrix}}
\def\ep{\end{pmatrix}}
\def\bea{\begin{eqnarray}}
\def\eea{\end{eqnarray}}
\def\nn{\nonumber}
\def\pref#1{(\ref{#1})}

\def\lb{\label}

\begin{document}

\title{Quantum Quench for inhomogeneous states in the non-local Luttinger model
}

\author{Vieri Mastropietro}
\affiliation{
Universit\'a di Milano, Via C. Saldini 50, 20133, Milano, Italy\\
email: Vieri.Mastropietro@unimi.it }

\author{ Zhituo Wang}
\affiliation{
Dipartimento di Matematica, Universit\`a di Roma Tre\\
Largo S. L. Murialdo 1, 00146 Roma, Italy\\
email: zhituo@mat.uniroma3.it}

\begin{abstract}
In the 
Luttinger model with non-local interaction we investigate, by exact analytical methods,  the time evolution of an inhomogeneous state
with a localized fermion 
added to the non interacting ground state. 
In absence of interaction
the averaged density has two peaks moving in opposite directions with constant velocities.  
If the state is evolved with the interacting Hamiltonian
two main effects appear. The first is that 
the peaks have velocities which are not constant but vary between a minimal and maximal value.
The second is that
a dynamical `Landau
quasi-particle weight' appears in the oscillating part of the averaged density, asymptotically vanishing with time, as consequence of the fact that fermions are not excitations of the 
interacting Hamiltonian. 
\end{abstract}


\maketitle

\subsection{Introduction}

Recent experiments on cold atoms \cite{B1} have motivated an increasing interest 
in the dynamical properties of many
body quantum systems which are {\it closed} and isolated from any
reservoir or environment \cite{P}.
Non equilibrium properties can be investigated by 
{\it quantum quenches},  in which
the system is prepared  
in a state and its subsequent time evolution driven by a many body Hamiltonian is observed. 
As the resulting dynamical behavior is the cumulative effect of the interactions between an infinite 
or very large number of particles,
the computation of local observables averaged over time-evolved states poses typically 
great analytical difficulties; 
the problem is then mainly studied in one dimension, see for instance
\cite{A},\cite{C},\cite{IC},\cite{IC1},\cite{IC2},\cite{IC3},\cite{L1},\cite{MG},\cite{Me},\cite{N},\cite{SM},\cite{B}.

A major difference with respect to the {\it equilibrium} case relies on the fact that in such a case
a form of
{\it universality} holds, ensuring that a number of properties are essentially {\it insensitive}
to the model details; for instance 
a large class of one dimensional system, named {\it Luttinger liquids} \cite{Ha}, have similar equilibrium properties irrespectively from the exact form of the Hamiltonian, and this fact
can be even proven rigorously under certain hypothesis using constructive RG methods \cite{BGM}.
Universality and independence from the details explain also why even crude approximations are able to capture the essential physics
of such systems. At non-equilibrium the behavior depends instead on model details; for instance 
integrability in spin chains dramatically affect the non equilibrium behavior
\cite{A1} while it does not alter the $T=0$ equilibrium properties \cite{M1}.
This extreme sensitivity to details or approximations asks for a certain number of analytical exact results
at non-equilibrium, to provide a benchmark for 
experiments or approximate computations.

One of the interacting fermionic system where non-equilibrium properties can be investigated
is the Luttinger model, which provides a great number of information in the equilibrium case. 
In this model the quadratic dispersion relation of non 
relativistic fermions is replaced with a linear dispersion relation, with the idea that
the properties are mainly determined
by the states close to the Fermi points, where the energy is essentially linear; a Dirac sea is introduced, filling all the states with negative energy. It is important to stress that there exist two versions of this model, 
the {\it local Luttinger model} (LNN)
and the {\it non local} Luttinger model (NLMM); in the former a local delta-like interaction is present while in the latter the interaction is short ranged but non local. 
At equilibrium such two models are often confused as they have similar behavior, due to the above mentioned insensitivity to model details; there is however no reason to expect that 
this is true also at non equilibrium. It should be also stressed that the LLM is
plagued by {\it ultraviolet divergences} typical of a  QFT and an {\it ad-hoc} regularization is necessary to get physical predictions; the short time or distance behavior depends on the chosen regularization.

The quantum quench of {\it homogeneous}
states in the LLM was derived in \cite{C},\cite{IC} and in the NLLM in \cite{Me}; the predictions agree for long times but are rather different for short times.
Regarding the quantum quench of {\it inhomogeneous} states, 
in \cite{L1} the dynamical evolution in the LLM of a 
{\it domain wall} state was considered, as an approximate description for the analogous problem in the spin XXZ spin chain.
It was found in \cite{L1} that the evolution in the free or interacting case is the same up to a finite renormalization of the parameters;
in particular
the front evolves with a {\it constant} velocity. Non constant velocities appear, from numerical simulations, 
in more realistic models like the XXZ chain \cite{SM}. 

In this paper we consider the evolution  of {\it inhomogeneous} states in the NLLM, using exact analytical methods in the infinite volume limit. In  particular we will consider the state obtained adding a particle to non interacting ground state or the vacuum.
In the absence of the interaction the particle moves with a ballistic motion
with a {\it constant} velocity, showing a typical "light cone"  dynamics. In presence of interaction, the dynamics is still ballistic (in agreement with the fact that the conductivity computed 
via Kubo formula is diverging), but the evolution is not simply the free one with a renormalized velocity; on the contrary, the evolution is driven by velocities which are
{\it non constant} and energy dependent.  Moreover the interaction produces a 
dynamical `Landau
quasi-particle weight' in the oscillating part, asymptotically vanishing with time; no vanishing weight is instead present in the non oscillating part. Note also that the expressions we get do not require any ultraviolet regularization, and correctly captures also the short time dynamics.

The plan of the paper is the following.
We introduce the NLLM in \S II  and in \S III we 
derive by this method
the ground state 2-point function and the average over an homogeneous quenched state. \S IV
contains our main result, namely 
the time evolution of an inhomogeneous state. In the Appendices the analytical derivation
of our results is exposed.

\subsection{The non-local Luttinger model}

The non-local Luttinger model (NLLM) Hamiltonian is \bea
&&H=\int_{-L/2}^{L/2} dxi ( :\psi^+_{x,1}\partial_x \psi^-_{x,1}:-
:\psi^+_{x,2}\partial_x \psi^-_{x,2}:) +\nn\\
&&\l \int_{-{L\over
2}}^{L\over 2} dx dy v(x-y) :\psi^+_{x,1} \psi^-_{x,1}:
:\psi^+_{y,2}\psi^-_{y,2}: \eea
where $\psi^\pm_{x,\o}=\frac{1}{ \sqrt{L}}\sum_k a_{k,\o} e^{\pm i
k x-0^+|k|}$, $\o=1,2$, $k=\frac{2\pi n}{ L}$ with $n\in N$ are fermionic
creation or annihilation operators, $v(x)$ is a {\it smooth short range
potential} such that
\be
|\hat v(p)|\le e^{-\k |p|}\ee and $::$ denotes Wick ordering. 
The main difference with the {\it local}
Luttinger model (LLM) is in the choice of the potential; in the LLM $v(x-y)=\d(x-y)$.
Usually in a low energy many body problem local or non short ranged interactions produce negligible differences; this is not true in the Luttinger model due to the linear relativistic dispersion relation, and ultraviolet divergences are produced in the LLM but not in the NLLM; as a consequence, the physical properties in the two models differ under several aspects.
We are choosing units so that $v_F=1$, where $v_F$ is the Fermi velocity.

The Hamiltonian can be
rewritten as 
\bea
&&H=H_0+V=\sum_{k>0} k[(a^+_{k,1} a^-_{k,1}+a^-_{-k,1}a^+_{-k,1})+\label{h}\\
&&( a^+_{-k,2} a^-_{-k,2}+a^-_{k,2}a^+_{k,2})]+\\
&&{2 \l\over L}\sum_{p>0}[\r_1(p)\r_{2}(-p)+\r_{1}(-p)\r_2(p)]+{\l\over L}\hat v(0)N_1 N_2\nn
\eea
where
\bea
&&\r_\o(p)=\sum_k a^+_{k+p,\o}a^-_{k,\o}\nn\\
&&N_\o=\sum_{k>0}(a^+_{k,\o}a^-_{k,\o}-a^-_{-k,\o}a^+_{-k,\o})\nn
\eea
The regularization implicit in the above expressions is that
$\r_\o(p)$ must be thought as $\lim_{\L\to\io }\sum_k
\chi_\L(k)\chi_\L(k+p) a^+_{k+p,\o}a^+_{k,\o}$ where $\chi_\L(k)$
is $1$ for $|k|\le\L$ and $0$ otherwise.
The Hamiltonian $H_\l$ as well as the $\r_\o(p)$ can be regarded
as operators acting on the Hilbert space $\HHH$ constructed 
applying
finitely many times creation or annihilation operators on
\be |0>=\prod_{k\le
0}a^+_{k,1}a^+_{-k,2}|vac>\label{gs} \ee
We define 
\bea\hat \psi^\pm_{\xx}&=&e^{i H_0 t}\psi^\pm_{\o,x} e^{-i H_0 t}\nn\\
&=&
{1\over\sqrt{L}}\sum_k a^\pm_{\o,k} e^{\pm i (kx -\e_\o kt)-0^+|k|}, \eea where $\e_1=+,\e_2=-$
so that 
\be
<0|\psi^\e_{\o,\xx}\psi^{-\e}_{\o,\yy}|0>=
{(2\pi)^{-1}\over i\e_\o(x-y)-i(t-s)+ 0^+}\ .
\ee
The basic property of the
Luttinger model is the validity of the following anomalous
commutation relations, first proved in \cite{ML}
\be
[\r_1(-p),\r_1(p')]=[\r_2(p),\r_2(-p')]={pL\over 2\pi}\d_{p,p'}\label{cr}\ee

Moreover one can verify that
\be\label{ss1} \r_2(p)|0>=0  \quad \r_1(-p)|0>=0. \ee
Other important commutation relations are the following
\bea
&&[H_0,\r_1(\pm p)]=\pm \r_1(\pm p)
\quad [H_0,\r_2(\pm p)]=\mp \r_2(\pm p)\nn\\
&&
[\r_\o,\psi^\pm_{\o,x}]=e^{ip x}\psi^\pm_{\o,x}\label{ss}
\eea
It is convenient, see \cite{ML}, to introduce an operator $ T=
{2\pi\over L}\sum_{p>0}[\r_1(p)\r_{1}(-p)+\r_{2}(-p)\r_2(p)]$ and
write $H_\l=(H_0-T)+(V+T)=H_1+H_2$. Note that $H_1$ commutes with
$\r_\o$ and $H_2$ can be written in diagonal form by the following
transformation
\bea&&\label{rho1} e^{iS} H_2 e^{-i S}=\tilde H_2\\
&&={2\pi\over L}\sum_p
\sech 2\phi_p[\r_1(p)\r_1(-p)+\r_2(-p)\r_2(p)]+E_0\nn\eea
so that
\be
 e^{iS}e^{iH_\l t} e^{-iS}=e^{i(H_0+D)t}
\ee
where
\be\label{S} S={2\pi\over L}
\sum_{p\not=0}\phi_p p^{-1}\r_1(p)\r_2(-p),\quad {\rm
tanh}\phi(p)=-{\l v(p)\over 2\pi}\ .\nn \ee
%
Defining $D=T+\tilde H_2$, we can write:
\be D={2\pi\over L}\sum_p  \s_p [\r_1(p)\r_1(-p)+\r_2(-p)\r_2(p)]+E_0\ee
where $\s_p=\sech 2\phi(p)-1$ and
and $[H_0,D]=0$.

The {\it ground state} of $H$ is
\be
|GS>=e^{iS}|0>\label{gs}\ee 
while $|0>$ is the ground
state of $H_0$. 

The relation between 
the creation or annihilation fermionic operators and the quasi-particle operators is can be defined as 
\be \psi_{x}=e^{i p_Fx}\psi_{x,1}+e^{-i
p_Fx}\psi_{x,2}\label{10}
\ee
and we call $e^{i p_Fx}\psi_{x,1}=
\tilde\psi_{x,1}$ and $e^{-i p_Fx}\psi_{x,-1}=\tilde\psi_{x,2}$.
where $p_F$ is the {\it Fermi momentum}. 
In momentum space this simply means that the momentum 
$k$
is measured from the Fermi points, that is $c_{k,\o}=\tilde c_{k+\e_\o p_F,\o}$, $\e_\o=\pm$.
Finally we recall that the XXZ spin chain model can be mapped in an interacting fermionc system; 
when the interaction in the third direction of the spin is missing ($XX$ chain) the mapping is over
a non interacting fermionic system with Fermi momentum $\cos p_F=h$. Therefore $|0>$ corresponds to the ground state of the $XX$ chain with magnetization $m$ such that $p_F=\pi({1\over 2}-m)$.

In the NLLM the
average of the 2-point function over the ground state is \cite{ML}, in the $L\to\io$ limit, see App. C
\bea
&&<GS|\psi^+_{\o,x}\psi^-_{\o,0}|GS>=
\frac{1}{2\pi}{1\over i \e_\o x+ 0^+}\label{ao}\\
&&\exp\int_0^\io dp
\frac{1}{p}\{2\sinh^2\phi_p (\cos p x-1)\}\nn
\eea
Asymptotically for large distances
\bea &&<GS|\psi^+_{\o,x}\psi^-_{\o,0}|GS> \sim O(|x|^{-1-\h}),\\
&&\h=\sinh^2\phi_0\nn\eea
implying that the average of the occupation number over the interacting ground state is $n_{k'+\e_\o p_F}\sim a+O(k^{'\h})$.

We now consider a quantum quench in which the interaction is switched on at $t=0$.
An interesitng quantity is the non interacting ground state evolved the the interacting Hamiltonian \cite{C}
\be < O_t|\psi^+_{\o,x}\psi^-_{\o,y}|O_t>=<0|e^{-i H t}\psi^+_{\o,x}\psi^-_{\o,y} e^{i H t}|0>
\label{asso1}
\ee
One finds, see App. D, in the limit $L\to\io$
\bea &&< O_t|\psi^+_{\o,x}\psi^-_{\o,0}|O_t>={1\over 2\pi}
{1\over i \e_\o x+ 0^+}
\label{a6}\\
&&
\exp\int_0^\io dp\frac{\g(p)}{p}\{
(\cos p x-1)(1-\cos 2p(\s_p+1)t)\},\nn
\eea
where $\g(p)=4\sinh^2\phi_p\cosh^2\ph_p$. Keeping $x$ fixed, see App. A
\bea \label{pazzz}&&\lim_{t\to\io} < O_t|\psi^+_{\o,x}\psi^-_{\o,0}|O_t> =
{1\over 2\pi}{1\over i \e_\o x+ 0^+}\nn\\
&&\exp\int_0^\infty\
dp\frac{1}{p}\{\g(p)(\cos\ p x-1)\}\label{a11} \eea
The 2-point function over $|0_t>$ reaches for $t\to\io$ a limit, similar but different 
with respect  to the average over the ground state 
\pref{ao} ; thermalization does not occur and memory of the initial state persists.
The difference between the limit of the quench and the ground state average is that the prefactor in the integrand (related to the critical exponent) 
is in one case
$\g(p)=4\sinh^2\phi_p\cosh^2\ph_p$
and in the other $2\sinh^2\phi_p$. 

The value of $< O_t|\psi^+_{\o,x}\psi^-_{\o,y}|O_t>$  in the LLM can be obtained from \pref{a11} 
replacing $\g(p),\s_p$ with $\g(0),\s_0$.
Doing that 
the integral in the exponent of \pref{ao}  becomes ultraviolet divergent and it requires a regularization; it is found ,see \cite{C}
\be {1\over 2\pi}{1\over i \e_\o x}
{1\over |x|^{\g(0)}}[\ 
{
x^2-v^2 t^2\over v^2 t^2}\ ]^{\g(0)\over 2}\label{io}\ee
where $v=1+\s_0$.
Comparing \pref{a6} with \pref{io} we see that the expressions in the LLM and the NLLM 
are rather different at short times; in the LLM there is a divergence at $t=0$ due to the ad hoc regularization which is of course absent in the NLLM.
The expressions qualitatively agree
if the limit $t\to\io$ is performed first but only if we consider the large distance
behavior; on the contrary for small distances the behavior is radically different. In the
NLLM one sees that the interaction has {\it no effect} at small distances (the integral in \pref{pazzz} is $=1$ as $x=0$);
this is what one expects in a solid state model,
as there are no high energy processes 
altering the short distances (or high momentum) behavior. On the contrary, 
from \pref{io}  we see that the interaction has a strong effect even for small $x$, as a singularity
$O(x^{-1-\g})$ is present,which is  a consequence
of the absence of an intrinsic cut-off in such a model.

\subsection{Quantum quench for the single particle state}

Let us consider now an {\it inhomogeneous} state obtained adding a particle to the non interacting ground state with Fermi momentum $p_F$; the case in which the particle is added to the vacuum is obtained
setting $p_F=0$. The state is the evolution  of $\psi^+_x|0>$ which by \pref{10} can be written as
\be |I_{\l,t}>=e^{i H_\l t} (e^{i p_F x} \psi^+_{1,x}+e^{-i p_F x}
\psi^+_{2,x})|0>. \ee
and we consider the average of the number operator $n(z)$
\be <I_{\l,t}|n(z)|I_{\l,t}> ,\ee
where $n(z)$ is the regularized version of the particle number $\psi^+_{z}\psi^-_z$, namely $n(z)=$
\be
{1\over 2}\sum_{\r=\pm}(\tilde \psi_{1, z+\r\e}^+ \tilde\psi_{2,z 
}^- +\tilde\psi_{2, z+\r\e}^+ \psi_{1, z}^- +\psi_{2, z+\r\e}^+ \tilde\psi_{2, z}^-  
+
\tilde\psi_{1, z+\r\e}^+ \tilde\psi_{1, z}^- )
\ee
One needs to
introducing a point spitting (the sum over $\r=\pm$) playing the same role as Wick ordering, and at the end the limit $\e\to 0$ is taken.
Note that
using the correspondence with the XXZ spin  modes, the state $|I_t>$ corresponds to adding
an excitation to the ground state of the $XX$ chain with total magnetization $m=1/2-p_F/\pi$.
It turns out that 
$<I_t|n(z)|I_t>$ is sum of several terms
\bea\label{wickord}
&&\langle 0|\tilde\psi_{1,
x}^- e^{i H t} \tilde \psi_{1, z+\r\e}^+ \tilde\psi_{2, z}^- e^{-i H t} \tilde\psi_{2, x}^+\ |0\rangle +\nn\\
&& \langle 0|\
\tilde\psi_{2, x}^-  e^{i Ht} \tilde\psi_{2, z+\r\e}^+
\psi_{1, z}^- e^{-i H t} \psi_{1, x}^+\  |0 \rangle+\label{real}\nn\\
&&\langle 0|\  \tilde\psi_{1, x}^-  e^{i H t} \tilde \psi_{2, z+\r\e}^+ \tilde\psi_{2, z}^- e^{-i Ht} \tilde\psi_{1, x}^+\  |0 \rangle + \nn\\
&&\langle 0|\ \tilde\psi_{2, x}^-  e^{i Ht}  \tilde\psi_{1, z+\r\e}^+ \tilde\psi_{1, z}^- e^{-i H t} \tilde\psi_{2, x}^+\  |0 \rangle+\label{nonrea2}\nn\\
&&\langle 0|\  \tilde\psi_{1, x}^-  e^{i H t} \tilde\psi_{1, z+\r\e}^+ \tilde\psi_{1, z}^- e^{-i H t} \tilde\psi_{1, x}^+
\  |0\rangle +\nn\\
&& \langle 0|\  \tilde\psi_{2, x}^-  e^{i H t}  \tilde\psi_{2, z+\r\e}^+ \tilde\psi_{2, z}^- e^{-i H t} \tilde\psi_{2, x}^+\  |0\rangle\label{nonreal3}.\nonumber
\eea
In the non-interacting case $\l=0$
the first term can be written as
\be
\langle0|\tilde\psi_{1,
x,t}^-  \tilde \psi_{1, z+\r\e}^+ \ |0 \rangle
\langle0|
\tilde\psi_{2, z}^- \tilde\psi_{2, x,t}^+\ |0 \rangle
\ee
so that in the limit $\e\to 0$ this term is equal to $e^{2ip_F(x-y)}(4\pi^2)^{-1}[(x-z)^2-t^2]^{-1}
$; a similar result is found for the second term. The third and fourth terms are vanishing as $\sum_{\r}{1\over \r\e}=0$; similarly the last two term
give $(4\pi^2)^{-1}[(x-z)\pm t]^{-2}$. Therefore
in the absence of interaction one gets 
\bea \label{finalbal} &&\lim_{L\to\io}<I_{0,t}|n(z)|I_{0,t}>=
\frac{1}{2\pi^2}{\cos 2p_F(x-y)
\over (x-z)^2-t^2}\nn\\
&&
+\frac{1}{4\pi^2}\ [{1\over(
(x-z)-t)^2}+{1\over( (x-z)+t)^2}]
\eea
The average of the density is sum of two terms, an oscillating and a non oscillating part (when the particle is added to the vacuum there are no oscillations $p_F=0$). At $t=0$ the density is peaked at $z=x$, where the average is singular. With the time increasing the particle peaks move in the left and right directions with 
constant velocity $v_F=1$ (ballistic motion); that is, the average of the density is singular at $z=x\pm  t$
and a "light cone dynamics" is found.

The interaction addresses in a quite non trivial way on the above dynamics. We get
 in the $L\to\io$ limit, see App. C
\bea \label{finalbala} &&\lim_{L\to\io}
<I_{\l,t}|n(z)|I_{\l,t}> =\nn\\
&&\frac{1}{4\pi^2}\ [{1\over(
(x-z)-t)^2}+{1\over( (x-z)+t)^2}]+\\
&&\frac{1}{4\pi^2}\ {e^{Z(t)}\over (x-z)^2-t^2}\nn\\
&&\big[\ e^{2i p_F(x-z)}
e^{Q_a(x,z,t)}+e^{-2i p_F(x-z)} e^{Q_b(x,z,t)}\ \big],
\nn \eea
where
\be Z(t)=\int_0^\io {dp\over p}\gamma(p)(\cos2p(\sigma+1)t-1)\ee
and $\gamma(p)=\frac{e^{4\phi(p)}-1}{2}$; moreover
$Q_a=$
\bea
&&\int_0^\io dp \frac{e^{-p 0^+}}{p}[(e^{ip(x-z)+ ip (\s_p+1)t}-e^{ip(x-z)+ ip t})\nn\\
&&+(e^{ip(x-z)- ip (\s_p+1)t}-e^{ip(x-z)- ipt})]\nn\eea 
and $Q_b=$
\bea
&&\int_0^\io dp \frac{e^{-p 0^+}}{p}[(e^{-ip(x-z)+ ip (\s_p+1)t}-e^{-ip(x-z)+ ip t})\nn\\
&+&(e^{-ip(x-z)- ip (\s_p+1)t}-e^{-ip(x-z)- ipt})]\big\}.\nn\eea

By looking at \pref{finalbala} we see first that the interaction
does not modify the non oscillating part. Regarding the oscillating part, it produces two main effects.
First of all the velocity of the peaks of is not anymore constant but {\it varies} between a maximal and minimal value. 
This is an effect which is absent in the LLM; indeed if we replace 
$\s_p$ with $\s_0$ we have
\be {1\over (x-z)^2-t^2}e^{Q_a}={1\over (x-z)^2-(1+\s_0)^2t^2},
\ee
so that is one gets the same expression as in the free case with a
renormalized velocity (a similar expression is valid for $Q_b$). 
The presence of non constant velocity is in agreement with the result of numerical simulations in the XXZ chain \cite{SM}.

The interaction has also another non trivial effect;  it introduces a dynamical "Landau
quasi-particle weight" in the oscillating part, asymptotically vanishing with time.
Indeed for large $t$ \be
\exp Z(t)= O(t^{-\g(0)}), \ee
while $Z(0)=1$. This vanishing weight can be physically interpreted as a consequence of the fact that fermions are not excitations of the interacting Hamiltonian. 
Finally note that the quasi-particle weight is $=1$ at $t=0$ and decreases at large $t$.

\subsection{Conclusions}

We have computed by exact analytical methods  the time evolution of an inhomogeneous state
with a localized fermion 
added to the non interacting ground state in the non local Luttinger model.
The interaction  does not produce a simple renormalization of the parameters
of the non interacting evolution; on the contrary it generates non constant velocities 
and 
a dynamical `Landau
quasi-particle weight' appears in the oscillating part of the averaged density, asymptotically vanishing with time. We believe that similar phenomena would be present also in the
evolution of
more complex initial states like a domain wall profile, and we plan to extend our methods to such a case.

\subsection{Acknowledgements}
We are very grateful to Alessandro Giuliani, Joel Lebowitz and Edwin Langmann for useful discussions and correspondences. We are grateful to the  ERC Starting Grant CoMBoS (grant agreement n.239694) for financial support.

\subsection{Appendix A}

In order to prove \pref{pazzz}
we set 
$z=p(1+\s)$
and we note that $\partial_p z(p)=H_p$ is bounded and different from zero: moreover $z$ is an increasing function of $p$
such that $p/z$ tends to a constant for $p\to 0$ and $p\to\io$.
Integrating by parts and using 
%
${ x
\sin p x\over p}\sim  x^2$, hence \pref{pazzz} follows.

In order to evaluate the large distance behavior of $Z(t)$ we use that
$\gamma(p)=
\frac{v(p)}{2\pi}$ and we write
$\int_0^{\infty} \frac{dp}{p}\gamma(p)(\cos 2\omega (p)pt - 1)$
as $\int_0^1+\int_1^\io$ where the second integral is bounded by a constant; in the first term we can write
$\g(p)=\g(0)e^{-\k p}+r(p)$ with $r(p)=o(p)$, and the integral containing $r(p)$ is again bounded by a constant. Note that
\be\gamma(0)\int_0^{1} \frac{e^{-\kappa p}dp}{p}(\cos 2\omega(0) pt - 1)=
\frac{\gamma(0)}{2}\log\frac{1}{\kappa^2+4\omega(0)^2t^2}
\ee
Moreover we can write $\o(p)=1+\s_p=\o(0)+f(p)$ with $f(p)=O(p)$ and
\bea
&&\int_0^{1} \frac{e^{-\kappa p}dp}{p}(\cos (2\omega (p)pt) -\cos 2\omega (0)pt)=\nn\\
&&\int_0^{1} \frac{e^{-\kappa p}dp}{p}(\cos (2\omega (0)pt)(\cos f(p) p t
-1)+\nn\\
&&\int_0^{1} \frac{e^{-\kappa p}dp}{p}\sin( 2\omega (0)pt)\sin (f(p) p t) \eea
Integrating by parts and dividing the integrals from $0$ to $t^{-1}$ ad from $t^{-1}$ to $1$
we get that both integrals are bounded by a constant.

\subsection{Appendix B}

In order to derive \pref{ao} we write $
<GS|\psi^+_{1,x}\psi^-_{1,y}|GS>$ as
\be
<0|e^{iS}\psi^+_{1,x}e^{-iS} e^{iS}\psi^-_{1,y}e^{-iS}|0>\label{ao1}
\ee
and
\be
 e^{i\e S}
\psi^-_{1, x} e^{-i \e S}= W_{1,x} R_{1,x}\psi^-_{1, x}
\ee
with $c(\phi)=\cosh\e\phi-1$, $s(\phi)=\sinh\e\phi$
\bea
&&W^\e_{1,x}=\nn\\
&&\exp\{-{2\pi\over L}\sum_{p>0}{e^{-0^+p}\over p}[\r_1(-p)e^{ipx}-\r_1(p)e^{-ipx}]c(\phi)\}\nn\\
&&R^\e_{1,x}=\nn\\
&&\exp\{-{2\pi\over L}\sum_{p>0}{e^{-0^+p}\over p}[\r_2(-p)e^{ipx}-\r_2(p)e^{-ipx}]s(\phi)\}\nn
\eea
so that \pref{ao1}
\be
<0|e^{iS}\psi^+_{1,x}W^{-1}_{1,x} R^{-1}_{1,x} R_{1,y} W_{1,y}\psi^-_{1,y}
|0>
\ee
By using the commutation relations  \pref{cr} and
$e^A e^B=e^B e^A e^{[A,B]}$ to carry $\r_1(p)$ ($\r_2(p)$) to the left (right) and $\r_1(-p)$ ($\r_2(-p)$) to the right (left) and using \pref{ss1}, we get \pref{ao}.

\subsection{Appendix C}

Let us consider now the interacting case starting from
\be 
\langle0|\psi_{1, x}^-  e^{i H t}  \psi_{1, z,}^+ \psi_{2, z}^- e^{-i H t} \psi_{2, x}^+\  |0 \rangle.
\ee
which can be rewritten as
\bea
&&\langle0|\psi_{1, x}^- e^{-iS} \{ e^{iS} e^{i H t}  e^{-iS}  e^{iS} \psi_{1, z}^+ 
e^{-iS}  e^{iS} e^{-i H  t}  e^{-iS}\}  \nn\\
&&\{e^{iS}  e^{i H t} e^{-iS}  e^{iS} \psi_{2, z}^-  e^{-iS}  e^{iS}e^{-i H t}  e^{-iS}\} e^{iS}
\psi_{2, x}^+\  |0 \rangle \nonumber
\eea

We use the relation
\be\label{lol}
e^{ i(H_0+D)t} e^{i S}
\psi^+_{1, x} e^{-i S}e^{-i(H_0+D)t}=
\bar\psi^+_{1, x,t}W^{-1}_{1,x,t} R^{-1}_{1,x,t}\nn
\ee
where $e^{ i(H_0+D)t} \psi^+_{1,x} e^{ -i(H_0+D)t}=\bar  \psi^+_{1,\xx}$
and, calling $c(\phi)=\cosh\phi-1$, $s(\phi)=\sinh\phi$
\bea
W_{1,x,t}&=&\exp\{-{2\pi\over L}\sum_{p>0}{1\over p}[\r_1(-p,t)e^{ipx}\nn\\
&-&\r_1(p,t)e^{-ipx}]c(\phi)\}\nn\\
R_{1,x,t}&=&\exp\{-{2\pi\over L}\sum_{p>0}{1\over p}[\r_2(-p,t)e^{ipx}\nn\\
&-&\r_2(p,t)e^{-ipx}]s(\phi)\}\nn
\eea
where $\r_1(\pm p,t)=e^{\pm i ( \s_p+1)t}\r_1(\pm p)  $, $\r_2(\pm p,t)=e^{\mp i (\s_p+1)t}\r_2(\pm p)$; 
moreover 
\be 
\bar\psi^\e_{1,\xx}=z_b \hat\psi^\e_{1,\xx} B_{1,+,\xx} B_{1,-\xx}=z_a B_{1,+,\xx} B_{1,-,\xx} \hat\psi^\e_{1,\xx}\label{lol1}
\ee
where 
$
B^\e_{1,+,\xx}=$
\be
\exp \e {2\pi\over L}\sum_{p>0} e^{-0^+ p}[\r_1( p)( e^{-i p x+i p(\s_p+1)t}-e^{-i p x+i  p t})]\nn
\ee
$
B^\e_{1,-,\xx}=$
\be
\exp -\e {2\pi\over L}\sum_{p>0}e^{-0^+ p}[\r_1(- p)( e^{i p x-i p(\s_p+1)t}-e^{i p x-i  p t})]\nn
\ee
and
$\hat\psi_{\xx,\o}^+=e^{iH_0 t}\psi_{x,\o}^+e^{-iH_0 t}$, $z_a=\exp \frac{2\pi}{L}\sum_p\frac{1}{p}(e^{ip\sigma_p t}-1)$
and $z_b=\exp \frac{2\pi}{L}\sum_p\frac{1}{p}(e^{-ip\sigma_p t}-1)$.

We write
\be
e^{-i S}\bar\psi^+_{1, x,t} W^{-1}_{1,x,t} R^{-1}_{1,x,t}
e^{iS}=e^{-i S}\bar\psi^+_{1, x,t}e^{iS}\bar
W_{y,t}^{-1} \bar R_{y,t}^{-1}
\ee
where $\bar
W_{y,t}, \bar R_{y,t}$ are equal to $
W_{y,t}, R_{y,t}$ in \pref{lol} with $\r(p)$ replaced by 
\bea\label{rho}
&& e^{-iS}\r_1(\pm p)e^{i S}=\r_1(\pm p)\cosh\phi(p)-\r_2(\pm p)\sinh\phi,\nn\\
&&e^{-iS}\r_2(p)e^{i S}=\r_2(\pm p)\cosh\phi(p)-\r_1(\pm p)\sinh\phi.
\eea
Note that $\bar
W_{1,y,t}\bar R_{1,y,t}=W^{-1}_{1,y} R_{1,y,0}$ so that $\bar
W_{1,y,0}\bar R_{1,y,0}(e^{-iS}\bar\psi^+_{1,x}e^{i S})=\bar\psi^+_{1,x}$.

It remains to evaluate $e^{-i S}\bar\psi_{1,x,t}e^{i S}$; we use \pref{lol1} so that it can be written as
\be
z_a \bar B^+_{1,+,x,t} \bar B^+_{1,-,x,t} (e^{-i S}
\hat\psi^+_{1,\xx}e^{i S})
\ee
where  $\bar B^\e_{1,+,x,t}, \bar B^\e_{1,-,x,t}$ are equal to $ B^\e_{1,+,x,t}, B^\e_{1,-,x,t}$
with $\r(p)$ replaced by \pref{rho}; moreover
\be
(e^{-i S}\hat\psi^+_{1,\xx}e^{iS})=\hat\psi^+_{1,\xx}W_{1,0,x,t} R^{-1}_{1,0,x,t} 
\ee
and $W_{1,0,x,t}$,  $R_{1,0,x,t}$ are equal to $W_{1,x,t}$,  $R_{1,x,t}$ with $\s_p=0$. In conclusion
\pref{asso1} is given by
which can be rewritten as
\bea
&&\langle0|\psi_{1, x}^- e^{-iS} \{ e^{iS} e^{i H t}  e^{-iS}  e^{iS} \psi_{1, z}^+ 
e^{-iS}  e^{iS} e^{-i H  t}  e^{-iS}\}  \nn\\
&&\{e^{iS}  e^{i H t} e^{-iS}  e^{iS} \psi_{2, z}^-  e^{-iS}  e^{iS}e^{-i H t}  e^{-iS}\} e^{iS}
\psi_{2, x}^+\  |0 \rangle \nonumber
\eea
and proceeding as above
\bea
&&\langle0|\psi_{1, x}(\bar B^+_{1,+,z,t} \bar B^+_{1,-,z,t}
\hat\psi_{z,t,1}^+
W_{1,0,z,t} R^{-1}_{1,0, z,t})\nn\\
&&\bar W^{-1}_{1,z,t}\bar R^{-1}_{1,z,t} \bar
W_{2,z,t}\bar R_{2,z,t}\nn\\
&&
(W_{2,0,z,t}^{-1} R_{2,0, z,t}  
\hat\psi^-_{z,t,2} \bar B^-_{2,+,z,t} \bar B^-_{2,-,z,t}
\psi_{2, x}^+)  |0 \rangle \nonumber
\eea
By using
\bea\label{timeshift1}
&&e^{\frac{2\pi}{L}\sum_p\frac{1}{p}F\rho_\o(\pm p)}\psi_{\o,x, t}\ e^{-\frac{2\pi}{L}\sum_p\frac{1}{p}F\rho_\o(\pm p)}\nn\\
&&=\psi_{\o, x,t} e^{-\frac{2\pi}{L}\sum_p\frac{1}{p}Fe^{\pm(ipx-ip t)}},\nonumber\\
&&e^{\frac{2\pi}{L}\sum_p\frac{1}{p}F\rho_\o(\pm p)}\psi_{\o, x, t}^\dagger\ e^{-\frac{2\pi}{L}\sum_p\frac{1}{p}F\rho_\o(\pm p)}\nn\\
&&=\psi^\dagger_{\o,x, t} e^{\frac{2\pi}{L}\sum_p\frac{1}{p}Fe^{\pm(ipx-ipt)}},
\eea
where $F$ is an arbitrary regular function,
and the Backer-Hausdorff formula to carry
$\r_1(p)$ ($\r_2(p)$) to the left (right) and $\r_1(-p)$ ($\r_2(-p)$) 
we get \pref{finalbala}.
\subsection{Appendix D}

We can write \pref{asso1} as
\bea\label{explct}
&&\langle 0|e^{-i S}\{ e^{iS}e^{iHt} e^{-iS} [e^{iS}\psi_{1,x}^+ e^{-iS}] e^{i S} e^{i H t}e^{-i S}\}e^{iS}\nn \times\\
&&\{e^{-i S} \{ e^{iS}e^{iHt} e^{-iS}[e^{i S}
\psi_{1,y}^- e^{-iS}] e^{iS}e^{-iH t} e^{-iS}\} e^{i S}|0\rangle\nn,
\eea
which is equal to
\bea
&&\langle 0|e^{-i S}e^{ i(H_0+D)t} [e^{iS}\psi_{1,x}^+ e^{-iS}] 
e^{-i(H_0+D) t} e^{i S}\nn\\ 
&&e^{-i S}e^{ i(H_0+D)t}[e^{i S}
\psi_{1,y}^- e^{-iS}]
e^{ -i(H_0+D)s} e^{iS}|0\rangle\nn
\eea
and by \pref{lol}
\be
\langle 0| e^{-i S}\bar\psi_{1,x,t}W^{-1}_{1,x,t}R^{-1}_{1,x,t}e^{iS}
e^{-i S}
W_{1,y,t}R_{1,y,t}\bar\psi_{1,x,t}e^{i S} |0\rangle\nn
\ee
from which we finally obtain  
\bea
&&\langle 0|\{e^{-i S}\bar\psi_{1,x,t}e^{i S }
\bar W^{-1}_{1,x,t}\bar R^{-1}_{1,x,t}\}\times\nn\\
&&\{\bar
W_{y,t}\bar R_{1,y,t}(e^{-iS}\bar\psi_{1,x,t}e^{i S}) |0\rangle\nn
\eea
As in the previous case, we now use the commutation relations  \pref{cr} and the relation
$e^A e^B=e^B e^A e^{[A,B]}$ to carry $\r_1(p)$ ($\r_2(p)$) to the left (right) and $\r_1(-p)$ ($\r_2(-p)$) to the right (left) and using \pref{ss1} and we get  \pref{ao}. The final expression coincide with the one found in \cite{Me} by a different method, namely using
a bosonization identity expressing the fermionc field in terms of bosons and Majorana operators.


\begin{thebibliography}{999999}

\bibitem{B1} I. Bloch, J. Dalibard and W. Zwerger, {\it Rev. Mod. Phys.}
80, 885 (2008).

\bibitem{P} Polkovnikov A, Sengupta, K, Silva A and Vengalattore M {\it Rev. Mod. Phys.} 83 863 (2011)

\bibitem{A} T. Antal, Z. Racz, G.M. Schutz,  {\em Phys. Rev. E} {\bf 59}, 5 4912 (1999)

\bibitem{C}
 M Cazalilla  {\em Phys.~Rev.~Lett.} {\bf 97} 156403 2006

\bibitem{IC}
Iucci A and Cazalilla MA 2009 {\em Phys.~Rev.} A {\bf 80} 063619

 \bibitem{IC1}
S.R. Manmana, S. Wessel, R.M. Noack, A. Muramatsu
{\it Phys. Rev. Lett}.98.210405 (2006)

\bibitem{IC2} P. Calabrese, J. Cardy
{\it Phys.Rev.Lett.} 96, 136801 (2006)

\bibitem{IC3} D.Fioretto, M. Mussardo
{\it  New J.Phys.}12:055015,2010 

\bibitem{L1} J. Lancaster A. Mitra {\it Phys. Rev. E} 81, 061134
(2010)

\bibitem{MG}
A. Mitra and T. Giamarchi, {\it Phys. Rev. Lett.} 107, 150602
(2011).

\bibitem{Me}C. Karrasch, J. Rentrop, D. Schuricht, V. Meden 
{\it Phys. Rev. Lett.} 109, 126406 (2012) 


\bibitem{N} W,Liu, N. Andrei {\it Phys. Rev. Lett} 2014

\bibitem{SM}
T. Sabetta, G. Misguich {\it Phys. Rev. B} 88, 245114 (2013)

\bibitem{B}
L. Bonnes, F. H. L. Essler, A. M. Läuchli {\em Phys. Rev. Lett.} 113, 187203 (2014)

\bibitem{ML} D. C. Mattis and E. H. Lieb  {\em J. Math. Phys.} {\bf 6}, 2304 (1965)

\bibitem{YY} C. N. Yang and C. P. Yang, Phys. Rev., 150, 321 (1966)


\bibitem{Ha} D. Haldane. {\em Phys. Rev. Lett.}
45, 1358 (1980)

\bibitem{BGM} G.Benfatto, P. Falco V. Mastropietro. {\em Phys. Rev. Lett.} 104 075701 (2010).

\bibitem{A1} J. Sirker, R.G. Pereira, I. Affleck  {\em Phys. Rev. B} 83, 035115 (2011) 

\bibitem{M1} V. Mastropietro. {\em Phys. Rev. E} 87, 042121 (2013) 




\end{thebibliography}
\end{document}